# Wavelength-Multiplexed 2D Beam Steering via a Passive Diffractive Network

Che-Yung Shen[1,2,3], Yuhang Li[1,2,3], Cagatay Isil[1,2,3], Tianyi Gan[1,3], Mona Jarrahi[1,3] and Aydogan Ozcan[1,2,3*]

[1]Electrical and Computer Engineering Department, University of California, Los Angeles, CA, 90095, USA

[2]Bioengineering Department, University of California, Los Angeles, CA, 90095, USA

[3]California NanoSystems Institute (CNSI), University of California, Los Angeles, CA, 90095, USA

[*]Correspondence to: ozcan@ucla.edu

## Abstract

Beam steering is traditionally achieved using mechanical scanners, liquid-crystal modulators, or optical phased arrays, which require active control and often impose trade-offs among speed, complexity, and footprint. Here, we introduce a wavelength-addressable diffractive optical network that transforms illumination wavelength into a high-dimensional control parameter for arbitrarily programmable 2D beam steering. The proposed passive architecture comprises cascaded spatially optimized diffractive layers, jointly designed using deep learning, to rapidly map distinct wavelengths to predefined/desired output angles. Unlike conventional single-layer dispersive optical elements, which are physically restricted to 1D linear mapping, this framework harnesses complex wavefront transformations to utilize the illumination wavelength as an intrinsic addressing key for arbitrary 2D beam steering, eliminating the need for mechanical scanning or electronic phase control. We numerically demonstrate wavelength-controlled beam steering across 625 wavelength channels spanning 400–750 nm, realizing a 25 × 25 array of independently addressable beam positions with subwavelength positioning accuracy and high channel fidelity. Unlike conventional gratings, which constrain wavelength routing to a linear trajectory, the proposed diffractive network performs nonlocal wavefront transformations, enabling arbitrary wavelength-to-angle mappings across a 2D field of view. We further validate the proposed framework experimentally in both the terahertz and visible spectral regimes, demonstrating wavelength-multiplexed beam steering using 3D fabricated passive diffractive layers at terahertz frequencies and phase-only spatial light modulators in the visible spectrum. This wavelength-addressable diffractive architecture establishes a compact and scalable paradigm for high-speed programmable beam steering, with potential applications in optical communications, routing, imaging, sensing, and emerging photonic information-processing systems.

## 1 INTRODUCTION

Beam steering is a foundational capability in modern optical and electromagnetic systems, underpinning applications ranging from wireless communications[1], light detection and ranging (LiDAR)[2] to high-resolution imaging and sensing[3]. Conventional beam steering technologies typically rely on, e.g., mechanical scanning systems[4], liquid crystal devices[5,6], micro-electromechanical systems (MEMS)[7,8] or optical phased arrays (OPAs)[9,10]. While these approaches have successfully enabled high-performance beam steering, they inherently present some limitations due to, e.g., mechanical inertia, limited steering speed and bulky system footprints. Furthermore, while simple passive dispersive elements such as traditional diffraction gratings can route light via wavelength tuning, they are constrained to a 1D linear trajectory. These challenges become particularly acute when scaling to high-density, multi-dimensional steering profiles over large operational regions.

Recent advances in deep learning–based inverse design of diffractive optical elements have introduced an alternative pathway for optical wavefront engineering. Prior work on multilayer diffractive optical networks/processors[11–19] demonstrated that spatially engineered dielectric layers with wavelength-scale features can be optimized through deep learning to manipulate optical fields with high precision. These diffractive processors are passive devices: once designed and fabricated, they perform the desired transformations of incident wavefronts without consuming active power. Such systems have been shown to enable quantitative phase imaging[20,21], optical phase conjugation and wavefront shaping[22], image denoising[23], sensing[24,25], nonlinear function approximation[26–28] and all-optical computation [29–32], reducing or even eliminating the need for digital post-processing for various applications.

Building on this paradigm, here we introduce a wavelength-multiplexed rapid 2D beam-steering system, comprising multiple spatially engineered, static dielectric layers that form a passive diffractive device (**Figure 1**). By jointly optimizing multiple diffractive layers through deep learning, the proposed passive processor learns arbitrary wavelength-dependent wavefront transformations that map individual wavelengths to independently specified beam directions across a 2D field of view. Unlike traditional beam-steering approaches that rely on dynamic phase tuning or mechanical motion, our method encodes beam-steering functionality directly into the static structure of the diffractive medium through its design. In this framework, each wavelength channel is deterministically routed to a specific steering angle, enabling wavelength-controlled beam direction without moving parts, electronic phase shifters, or active modulation. This architecture effectively transforms spectral diversity into a high-dimensional control space for beam routing. Distinct from a single-layer grating, which maps wavelengths strictly along a 1D dispersive trajectory, our deep learning-optimized multilayer diffractive structure enables complex, non-local wavefront transformations. By introducing cascaded, spatially engineered phase modulation boundaries, the diffractive network breaks free from the spatial restrictions of linear dispersion. The high-dimensional design parameters within the cascaded dielectric layers suppress inter-channel spectral crosstalk and optimize angular selectivity, enabling arbitrary routing of light independently across a full 2D field of view.

We numerically demonstrated large-scale wavelength-controlled beam steering over 625 distinct wavelength channels spanning 400–750 nm, corresponding to a 25 × 25 array of output beam positions. We further experimentally validated the proposed concept in both the terahertz and visible spectral regimes, demonstrating wavelength-multiplexed beam steering using fabricated passive diffractive layers at terahertz wavelengths and phase-only spatial light modulators (SLMs) in the visible spectrum. By combining deep learning–based inverse design with passive diffractive optics, this work establishes a new method toward scalable, multiplexed, and energy-efficient beam steering. This capability emerges from the large number of trainable spatial degrees of freedom distributed across multiple diffractive layers, which collectively realize wavelength-dependent transformations that are not constrained by the local phase-gradient relationship governing conventional dispersive optics. Crucially, owing to the wavelength-scalability of diffractive optical networks, the same operational principle can be applied across different regions of the electromagnetic spectrum by appropriately scaling the diffractive features and adapting the fabrication approach to the target spectral band.

## 2 Results

**Figure 1** illustrates a schematic of our diffractive beam steering framework. The system is sequentially illuminated by spatially coherent light with varying wavelengths from the set $\{\lambda_1, \lambda_2, \ldots, \lambda_{N_w}\}$, ordered from the longest to the shortest wavelength, where each wavelength corresponds to a distinct output beam position. Following the input aperture, the diffractive beam steering design consists of multiple modulation layers made of passive dielectric materials. Each layer contains the same number of diffractive features with a lateral size of $\sim\lambda_{N_w}/2$ and a trainable/learnable thickness, providing a phase modulation range spanning 0-2π for all wavelengths within the input illumination band. These layers, including the input/output planes, are optically connected to each other through free-space diffraction. The uniform input optical fields $\{\boldsymbol{i}_w\}$ ($w \in \{1,2,\ldots,N_w\}$) are instantaneously processed by the diffractive beam steering model to produce output complex fields $\{\boldsymbol{o}_w\}$ for each wavelength $\lambda_w$. The resulting intensity distributions yield the output intensity patterns $\boldsymbol{O}_w$ at each one of the $N_w$ wavelengths. During the design process, each wavelength channel is assigned to a predefined point-like target image in the output plane, enabling $N_w$ distinct output beam positions/angles across the respective wavelength channels. The diffractive design is optimized using error backpropagation and stochastic gradient descent [25] to minimize a custom loss function $\mathcal{L}$, defined based on the mean-squared error (MSE) between the output intensity profiles and their corresponding ground truths across all wavelengths (refer to the **Experimental Section** for details).

To demonstrate the proof-of-concept of our beam steering design, we encoded 625 desired beam positions in 2D and assigned them to distinct wavelength channels, covering an output field of view (FOV) of 25 × 25 points using 625 wavelength channels within a range of 400 nm to 750 nm. **Figure 2a** illustrates the resulting output performance of this beam steering design in terms of peak signal-to-noise ratio (PSNR) values, where the blue curve represents the performance of the output beam profiles at the assigned wavelengths, yielding an average PSNR of 33.32 ± 0.41 dB. The orange curve corresponds to the output performance at shifted illumination wavelengths (i.e., $\lambda_w + \Delta\lambda_w/2$), which were *not* included in the training stage but lie between the assigned/pre-determined wavelengths; at these intermediate, shifted illumination wavelengths, we still observe a decent performance with an average output PSNR of 32.48 ± 0.42 dB. These results demonstrate that the diffractive beam steering design can effectively project light even at intermediate illumination wavelengths between the pre-trained points, without explicit training at those wavelengths. Additionally, we observed periodic performance drops in the PSNR curves for both the assigned wavelengths and shifted wavelengths, primarily caused by the 2D zigzag scanning pattern, where beam positions located near the turning points of the 2D scanning trajectory introduce slightly larger errors. These localized performance drops occur because the sharp spatial-frequency transitions required at the boundaries of the 2D FOV demand higher angular scattering.

The accuracy of the output beam positioning is further supported by the lateral center of mass (COM) error metrics reported in **Figure 2b**, displaying a subwavelength average error of ~0.18 $\lambda_\mathrm{m}$ for the assigned/pre-determined illumination wavelengths and ~1.16 $\lambda_\mathrm{m}$ for external/shifted wavelengths. Furthermore, **Figure 2c** depicts the output beam profiles, which closely match the ground truth, even for shifted illumination wavelengths (external generalization), thereby demonstrating precise, programmable beam steering capability by varying the illumination wavelength. Additionally, **Video S1 (Supporting Information)** shows a movie of continuous-beam steering as the illumination wavelength is varied from 400 to 750 nm. This diffractive beam steering design also achieved an angular steering resolution $\Delta\theta$ of ~0.11°, defined by the angular spacing between adjacent wavelength-addressed beam positions in the output FOV. To further quantify the performance of this diffractive approach, we define a spectral beam-steering figure of merit, i.e., $FOM_{steering}= \Delta\theta/\Delta\lambda$, where $\Delta\lambda$ is the wavelength spacing between adjacent channels, which achieves $FOM_{steering} \approx 0.196°/nm$ for the presented diffractive device. These numerical results highlight the capabilities of our diffractive optical networks to dynamically adjust the beam position across the output FOV with sub-wavelength lateral positioning accuracy by varying the illumination wavelength, demonstrating their potential for efficient, programmable beam steering solutions.

We also conducted experimental demonstrations of our diffractive beam-steering system in the terahertz and visible parts of the electromagnetic spectrum. First, using the same diffractive architecture and training techniques as the numerical model depicted in **Figure 2**, we developed a wavelength-multiplexed diffractive beam-steering design that operates across five wavelengths: 0.64 mm, 0.68 mm, 0.72 mm, 0.76 mm, and 0.8 mm, as illustrated in **Figure 3a**. In our experimental set-up, we utilized a diffractive beam-steering system comprising two dielectric diffractive layers ($L_1$ and $L_2$). This diffractive design effectively projects beam patterns at the output plane, assigning a unique beam pattern to each illumination wavelength. For the optimization of our experimental designs, we generated a set of target images for deep learning-based training of the diffractive layers, where each image contained a single illuminated pixel (i.e., the desired beam position), and the others remained off. Both the numerical design and the 3D fabricated masks are shown in **Figure 3b**. Following the 3D assembly and alignment of these fabricated diffractive layers, we employed a tunable terahertz source coupled with a detector to capture the intensity distribution at the output plane as a function of the illumination wavelength. Detailed schematics and images of this experimental set-up are shown in **Figure 3c,d**. **Figure 3e** shows the experimental measurements, confirming the alignment of the output images with the numerically simulated patterns and the target ground truth images. The output THz beam was accurately steered to the designated locations within the output FOV, consistently producing the desired beam spot at the assigned spatial position for each illumination wavelength. These experimental results affirm the diffractive model's capability for effective wavelength-multiplexed beam steering, consistent with our earlier numerical results and analyses shown in **Figure 2**.

To further validate the applicability of the proposed architecture across different spectral regimes, we experimentally demonstrated wavelength-multiplexed diffractive beam steering in the visible spectrum. Using the same device architecture and training procedure as the numerical model shown in **Figure 2**, we designed a wavelength-controlled diffractive beam steering system that routes incident light to distinct output positions across nine illumination wavelengths: 710, 680, 650, 620, 590, 560, 530, 500 and 470 nm. For this proof-of-concept experiment, the desired steering positions were defined on a 3 × 3 output grid, where each wavelength channel was assigned to a unique beam location within the output field of view. The experimental set-up, shown in **Figure 4a**, consisted of a multispectral illumination source, two cascaded diffractive layers, $L_1$ and $L_2$, implemented using two phase-only SLMs, and a complementary metal-oxide-semiconductor (CMOS) image sensor to record the wavelength-dependent output intensity distributions. A photograph of the experimental set-up is provided in **Figure 4b**, and the optimized diffractive phase profiles displayed on the SLMs are shown in **Figure 4c**.

To mitigate the impact of experimental misalignments, calibration errors, and other physical imperfections that arise when transferring digitally optimized diffractive designs to an optical hardware system, we employed an in-situ learning strategy[33,34] in the visible light experiment (see the **Experimental Section**). As in-situ training progressed, the loss value decreased, while the PCC and SNR metrics increased, indicating improved agreement between the experimentally measured beam profiles and their assigned target positions, as shown in **Figure 4d**. After sequentially optimizing the diffractive layers across all the visible wavelength channels, the final phase profiles were fixed and tested under each illumination wavelength. **Figure 4e** presents the experimentally measured beam steering results after the in-situ optimization, together with the corresponding numerical simulations and target beam positions. The measured output beams were correctly steered to their designated spatial locations with good agreement with the simulations and targets. The visible-light beam steering system achieved an angular steering resolution of ~0.06°, corresponding to the angular spacing between adjacent beam positions in the 3 × 3 output grid. This performance was achieved through wavelength-controlled diffractive routing without mechanical scanning or active electronic phase tuning, highlighting the potential of the proposed architecture for compact, passive, and spectrally multiplexed beam steering in the visible spectrum.

## 3 Discussion

We introduced a passive optical processor that uses wavelength as an addressing dimension to achieve arbitrary 2D beam steering beyond the fundamental routing limitations of conventional dispersive optics. For future practical implementations, several fabrication and system-level considerations must be addressed. A key challenge is the sensitivity of multilayer diffractive systems to interlayer misalignment. Lateral or axial shifts between diffractive layers may distort the intended wavefront transformation and degrade the beam steering performance. This issue can be mitigated by incorporating random lateral and axial shifts between the diffractive layers during training to improve the robustness of the diffractive design [17,35,36]. From a hardware perspective, alignment errors can be further reduced through precision alignment stages, alignment marks, monolithic 3D fabrication approaches (using, e.g., two-photon polymerization[37]), or spacer-defined self-alignment techniques. In addition, as demonstrated in our visible light experiment in **Figure 4d**, in-situ learning provides an effective way to compensate for residual misalignments, calibration errors, and other hardware imperfections after 3D system assembly. By optimizing the diffractive response directly on the physical optical hardware, in-situ training improved the agreement between the measured beam profiles and their target positions, leading to enhanced beam steering performance. However, this strategy is most readily applicable to programmable diffractive implementations, such as SLM-based systems. For fully fabricated passive diffractive structures, the phase profiles are fixed after fabrication; therefore, in-situ learning cannot directly modify the physical structure. In such cases, fabrication-aware and misalignment-robust inverse design, accurate assembly, and post-fabrication calibration become particularly important.

Another practical concern arises from finite phase quantization (i.e., phase bit depth) imposed by fabrication constraints, especially in visible-wavelength implementations where high-resolution nanofabrication is required[38]. Limited phase resolution can reduce beam steering efficiency and introduce undesired sidelobes in the output beam profile. To address this issue, quantization-aware optimization can be incorporated during the inverse design stage, allowing the diffractive layers to be trained under discrete phase levels (e.g., 2–8 bits). By integrating fabrication-aware optimization, diffractive beam steering systems can achieve high robustness and reliability in real-world implementations.

Notably, the trade-off between beam steering performance and diffraction efficiency is also critical for the practical implementation of diffractive beam steering systems. Optimizing solely for beam-profile fidelity may yield accurate beam locations at the output FOV but with reduced diffraction efficiency, whereas overly emphasizing output efficiency may compromise beam localization, angular selectivity, or inter-channel crosstalk suppression. This trade-off can be mitigated by incorporating efficiency-aware terms into the training objective. In this work, we explicitly accounted for this balance by including a diffraction-efficiency penalty in the overall loss function for the experimentally validated diffractive beam-steering designs shown in **Figure 3a** and **Figure 4a** (refer to the **Experimental section** for details). This formulation guides the model to maintain the desired output diffraction efficiency while preserving accurate wavelength-multiplexed beam routing.

Because the presented framework maps specific wavelengths directly to discrete diffraction angles, the system's operational resolution is coupled to the spectral linewidth of the illumination source. A laser with a large linewidth will introduce spatial blurring at the output plane due to localized angular dispersion. Based on our calculated spectral figure of merit ($FOM_{steering}$), an illumination source linewidth of $< 1$ nm is ideal to ensure clear spatial isolation between adjacent spots in high-density beam steering channels, whereas broader sources can be utilized for lower-density beam routing tasks.

In summary, we developed a wavelength-multiplexed diffractive beam steering system capable of routing multiple wavelength channels into predefined angular directions using cascaded spatially optimized dielectric layers that are static. The deep learning–based inverse design framework effectively suppresses spectral crosstalk and optimizes angular selectivity, ensuring high steering fidelity across all wavelength channels. Compared with other beam-steering platforms, which typically rely on, e.g., actively addressed phase pixels, mechanical scanning or dense waveguide networks, our approach encodes the steering function into a passive, wavelength-multiplexed diffractive processor, enabling <0.1° angular steering resolution through spectral selection alone. Furthermore, by combining wavelength multiplexing with additional degrees of freedom such as phase and/or polarization multiplexing[39–42], the beam steering capacity and functionality of diffractive optical processors can be further expanded. Importantly, owing to the scale-invariant nature of diffractive optics, the presented architecture can be translated across different regions of the electromagnetic spectrum—from visible to terahertz frequencies—by scaling each diffractive design proportional to the wavelength, i.e., without the need to redesign the diffractive layers, enabling flexible implementation using a wide range of fabrication techniques tailored to different spectral bands. These capabilities highlight the potential of wavelength-multiplexed diffractive beam steering as a scalable platform for next-generation LiDAR, optical communication, sensing, and imaging technologies.

## 4 Experimental Section

### Optical forward model of the wavelength-multiplexed diffractive beam steering design

Our diffractive beam steering system consists of $K$ cascaded diffractive layers, each comprising thousands of independent and spatially distributed subwavelength features. In the numerical model, each layer is approximated as a thin planar modulation surface that applies a complex-valued transmission function to an incident coherent optical field. For the $s^{\text{th}}$ diffractive feature on the $l^{\text{th}}$ layer located at $(x_s, y_s, z_l)$, the complex transmission coefficient at wavelength $\lambda$ is determined by its material thickness $h_s^l$, which can be expressed by the following equation:

$$t(x_s, y_s, z_l; \lambda) = \exp\left(\frac{-2\pi\kappa(\lambda)h_s^l}{\lambda}\right)\exp\left(\frac{-j2\pi(n(\lambda) - n_{\text{air}})h_s^l}{\lambda}\right) \quad (1)$$

Here, $n(\lambda)$ and $\kappa(\lambda)$ denote the real and imaginary parts of the material's complex refractive index $\tilde{n}(\lambda)$, i.e., $\tilde{n}(\lambda) = n(\lambda) + j\kappa(\lambda)$. For all the numerical simulations of diffractive beam steering designs within the visible spectrum reported in this paper, we selected N-BK7 as the material of the diffractive layers[43]. N-BK7 exhibits negligible absorption in the visible range, and therefore $\kappa(\lambda)$ was assumed to be 0. For the experimentally tested diffractive beam steering model in the terahertz spectrum, $n(\lambda)$ and $\kappa(\lambda)$ were set based on the measurements from a terahertz spectroscopy system[35].

Each diffractive feature thickness is parameterized as:

$$h = h_{\text{learnable}} + h_{\text{base}} \quad (2)$$

where $h_{\text{learnable}} \in [0, h_{\max}]$ is optimized during training, and $h_{\text{base}}$ represents a fixed substrate thickness. For the numerical simulated designs within the visible spectrum, $h_{\max}$ was set as 664.5 nm, corresponding to a full phase modulation range from 0 to $2\pi$ for the longest wavelength ($\lambda_1$). $h_{\text{base}}$ was empirically chosen as 700 nm. For the experimentally validated diffractive models operated within the terahertz spectrum, $h_{\max}$ was set as 1.4 mm, while the base thickness was set as 0.2 mm. For the experimentally validated visible light diffractive beam steering system in **Figure 4a**, as we utilized phase-only SLMs, we applied a similar model and treated the diffractive layer as a pure phase modulator, with a trainable structured reflection coefficient of:

$$r(x_s, y_s, z_l; \lambda) = \exp(-j\phi(x_s, y_s; \lambda)) \tag{3}$$

Here, $\phi(x_s, y_s; \lambda)$ is the phase modulation value for the corresponding diffractive feature, optimized during the training process.

Free-space propagation between adjacent diffractive layers was modeled using Rayleigh–Sommerfeld scalar diffraction theory. In the simulation, the diffraction process is formulated as a linear, shift-invariant operator with an impulse response. Each $s^{\text{th}}$ diffractive feature on the $l^{\text{th}}$ layer at $(x_s, y_s, z_l)$ acts as the source of a secondary wave, generating a complex field at wavelength $\lambda$ given by the equation:

$$w_s^l(x, y, z; \lambda) = \frac{z - z_l}{\left(r_s^l\right)^2}\left(\frac{1}{2\pi r_s^l} + \frac{n}{j\lambda}\right)\exp\left(\frac{j2\pi n r_s^l}{\lambda}\right) \tag{4}$$

where $r_s^l = \sqrt{(x - x_s)^2 + (y - y_s)^2 + (z - z_l)^2}$. These secondary propagate to the next layer (the $(l+1)^{\text{th}}$ layer), where they are spatially superimposed. Therefore, the optical field that reaches the $p^{\text{th}}$ diffractive feature in the $(l+1)^{\text{th}}$ layer, located at $(x_p, y_p, z_{l+1})$, can be computed by the convolution of the complex amplitude $u_s^l$ from the previous layer with the impulse response function $w_s^l(x_p, y_p, z_{l+1}; \lambda)$. The resulting field is then modulated by the transmission function $t(x_p, y_p, z_{l+1}; \lambda)$ of the $(l+1)^{\text{th}}$ diffractive layer, which can be expressed as:

$$u_p^{l+1}(x, y, z; \lambda) = t(x_p, y_p, z_p; \lambda)\sum_s u_s^l w_s^l(x_p, y_p, z_{l+1}; \lambda) \tag{5}$$

All numerical simulations in the terahertz spectrum for the diffractive beam steering designs used a lateral sampling interval equal to half of the shortest wavelength, which also defines the lateral feature size. As for the experimentally validated diffractive beam steering design in the visible spectrum, as shown in **Figure 4a**, the spatial sampling period was set to be 2 μm.

For the K=8 diffractive beam steering design shown in **Figure 2**, which operates at $N_w$ = 625, the sizes of the input aperture and the output FOV were both set to be ~$69.56\lambda_{\text{m}} \times 69.56\lambda_{\text{m}}$. The output field of view consists of $N_x \times N_y$ = 25 × 25 pixels, resulting in each output pixel having a size of ~$2.78\lambda_{\text{m}} \times 2.78\lambda_{\text{m}}$. To achieve the beam steering task, we designed the diffractive model to possess 800 × 800 diffractive features per layer, resulting in each layer a diffractive area of ~$278\lambda_{\text{m}} \times 278\lambda_{\text{m}}$. For the experimentally validated diffractive beam steering design in the terahertz spectrum shown in **Figure 3a**, the output FOV was set as ~$30\lambda_{\text{m}} \times 30\lambda_{\text{m}}$, which is divided into 3 × 3 pixels, where each pixel has a size of ~$10\lambda_m \times 10\lambda_m$. The operational wavelengths for the experimental THz validation were selected in a range of [0.64, 0.68, 0.72, 0.76, 0.8] mm.

For the experimentally validated visible-light diffractive beam-steering system shown in **Figure 4a**, the output FOV was set to 0.675 mm × 0.675 mm and discretized into a 3 × 3 grid of target beam positions. Each output grid cell had a lateral size of 0.225 mm × 0.225 mm. The illumination wavelengths used for the experimental validation were selected as [710, 680, 650, 620, 590, 560, 530, 500, 470] nm. We designed the diffractive model to possess 500 × 500 diffractive features per layer. Based on the SLM pixel sizes used in the experiment, the corresponding active diffractive area is 4 mm × 4 mm for the first layer (pixel size $s_1$=8 µm) and 2.25 mm × 2.25 mm for the second layer (pixel size $s_2$=4.5 µm).

**Data preparation and other implementation details**

To train and evaluate our diffractive beam-steering designs, we developed a custom dataset of 625 images specifically designed to optimize beam steering. Each image in this dataset is structured as a 25 × 25 grid, with only a single pixel activated, representing a distinct beam steering direction. Through the dataset, the activated pixel systematically shifts from the top-left to the bottom-right corner of the grid following a zigzag trajectory, effectively covering the entire FOV. This comprehensive coverage ensures that the diffractive model learns to accurately steer beams in all directions within the defined spatial resolution, covering the entire FOV. For training the experimentally validated diffractive beam steering designs, we designed a custom dataset comprising 9 images, each with a 3 × 3 grid, with only a single pixel activated, representing the beam steering direction.

All numerical simulations and training procedures for our diffractive beam-steering designs were implemented in Python (version 3.10.4) and JAX (version 0.4.1). We employed the Adam optimizer, utilizing the default parameters provided by the OPTAX library, with a learning rate set at 0.001 and a batch size of 128. The diffractive models were trained for 200 epochs on a workstation equipped with an Nvidia GeForce RTX 3090 GPU, an Intel Core i9-11900 CPU and 128 GB of RAM. The training of the 8-layer diffractive beam steering design shown in **Figure 2** took ~16 days to complete. It is important to emphasize that this training process for our diffractive beam-steering design on a computer is a one-time effort.

**Details of the experimental diffractive beam steering systems**

We experimentally validated the proposed wavelength-multiplexed diffractive beam-steering framework using two different optical set-ups: a terahertz system based on 3D-fabricated passive diffractive layers and a visible-light system implemented with phase-only SLMs.

The diffractive beam-steering design shown in **Figure 3d** was experimentally evaluated using a continuous-wave (CW) terahertz system. The set-up included a terahertz source consisting of a Virginia Diode Inc. WR9.0M SGX/WR4.3x2 WR2.2 modular amplifier/multiplier chain (AMC), coupled with a diagonal horn antenna (Virginia Diode Inc. WR2.2). At the input of the AMC, a 10-dBm radiofrequency (RF) signal was provided at frequencies of 13.03, 12.25, 11.56, 10.97 or 10.42 GHz (fRF1), which was subsequently multiplied 36 times, generating CW radiation at frequencies of 0.469, 0.441, 0.416, 0.395 or 0.375 THz, corresponding to illumination wavelengths of 0.64, 0.68, 0.72, 0.76 or 0.8 mm, respectively. The AMC output was modulated with a 1-kHz square wave for lock-in detection. The system utilized an XY positioning stage composed of two Thorlabs NRT100 motorized stages to move a single-pixel mixer (Virginia Diode Inc. WRI 2.2), which performed a 2D scan of the output intensity distribution, with a step size of 0.8 mm. For local oscillation, the detector was fed a 10-dBm RF signal at the same frequencies (13.03, 12.25, 11.56, 10.97, or 10.42 GHz, fRF2), which down-converted the terahertz signal to 1 GHz. This down-converted signal was then amplified by a low-noise amplifier (80 dBm gain) and passed through a KL Electronics 3C40-1000/T10-O/O bandpass filter centered at 1 GHz (+/- 10 MHz) to suppress noise from unwanted frequency bands. After linear calibration using an HP 8495B tunable attenuator, the signal was directed to a Mini-Circuits ZX47-60 low-noise power detector. The detector output voltage was processed by a Stanford Research SR830 lock-in amplifier, using a 1-kHz square-wave modulation signal as a reference for calibration on a linear scale.

For the fabrication of the diffractive beam-steering design depicted in Figure 3a, an Objet30 Pro 3D printer by Stratasys was used to print the diffractive layers. To ensure alignment with our optical forward model for the experimental diffractive design, a 3D-printed holder was fabricated using the same printer. This holder enabled precise positioning of the printed diffractive layers, ensuring their accurate 3D assembly.

For the visible light diffractive beam steering experiment shown in **Figure 4a**, a tunable laser source (WhiteLase-Micro; Fianium Ltd, Southampton, UK) provided illumination across different wavelength channels, with each spectral channel filtered to a ~5 nm bandwidth using an acousto-optic tunable filter. Two phase-only liquid-crystal-on-silicon (LCOS) SLMs were used as diffractive layers: a HOLOEYE PLUTO-2.1 device (pixel size $s_1$=8 μm, 1920 × 1080) and a HOLOEYE LUNA device (pixel size $s_2$=4.5 μm, 1920 × 1080). In the experiment, the central (500 × 500) pixels of each SLM were used as the trainable diffractive aperture. The output intensity distributions were recorded using a CMOS camera (Basler ace acA1920-40gm). The propagation distance between the two SLM planes was 3 cm, followed by an additional propagation distance of 17.6 cm from the second SLM to the camera sensor.

The diffractive layers were first optimized in silico. For each wavelength channel $\lambda_w$, the wavelength-dependent phase values of the diffractive layers, denoted as $\phi(\lambda_w)$, were optimized to minimize the loss functions defined in Equation (10). However, the numerical forward model cannot fully account for uncontrolled experimental factors, including optical misalignments, calibration errors, aberrations, system noise, and other hardware imperfections. To address these challenges, we further refined the diffractive layers directly on the optical set-up using a model-free in-situ training strategy based on Proximal Policy Optimization (PPO)[33,34]. In this framework, rather than directly optimizing a deterministic phase profile, we optimized a parameterized probability distribution $\pi(\phi(\lambda_w)|\theta)$ with parameters $\theta$, from which the phase values $\phi(\lambda_w)$ are sampled. This enabled the system to learn from optical measurements acquired directly from the physical hardware. Therefore, we formulated an in-situ learning loss function for each wavelength channel $\lambda_w$ as:

$$\mathcal{L}_w(\theta) = \mathbb{E}_{\pi(\phi(\lambda_w)|\theta)}[\mathcal{L}_w(\phi(\lambda_w))] = \int \mathcal{L}_w(\phi(\lambda_w))\pi(\phi(\lambda_w)|\theta)d\phi(\lambda_w)$$

Using Monte Carlo integration with $M$ samples and defining the advantage function as $A_{w,j} = -\mathcal{L}_w(\phi_j(\lambda_w))$, this loss function can be estimated as:

$$\mathcal{L}_w = -\frac{1}{M}\sum_{j=1}^{M} A_{w,j}\pi\left(\phi_j(\lambda_w)|\theta\right)$$

For the PPO-based in-situ training, we used the clipped surrogate objective:

$$\mathcal{L}_w^{PPO}(\theta) = -\mathbb{E}\left[\min(r(\theta)A'_{w,j}, \mathrm{clip}(r(\theta), 1-\epsilon, 1+\epsilon)A'_{w,j})\right]$$

where $A'_{w,j}$ is the normalized advantage function defined by $A'_{w,j} = \frac{A_{w,j}-\mu_A}{\sigma_A}$. Here, $\mu_A$ and $\sigma_A$ are the mean and standard deviation of $\{A_{w,1}, A_{w,2}, \dots, A_{w,M}\}$. The probability ratio $r(\theta_w) = \frac{\pi(\phi(\lambda_w)|\theta)}{\pi(\phi(\lambda_w)|\theta_{old})}$ measures the deviation between the updated and previous policies, and $\epsilon$ is a hyperparameter that prevents excessive parameter updates. $\epsilon$ was empirically chosen as 0.2.

After collecting a batch of $M$ optical measurements corresponding to the parameters drawn from the current policy distribution $\pi(\phi(\lambda_w)|\theta)$, we performed $X$ iterations of updates without requiring additional physical measurements. During these iterations, the phase values remain constant on the diffractive hardware while only updating the distribution $\pi(\phi(\lambda_w)|\theta)$ digitally. Specifically, we set the policy $\pi(\phi(\lambda_w)|\theta)$ as a Gaussian $\mathcal{N}(\mu, \sigma^2)$, with the mean $\mu$ being the trainable parameter and $\sigma$ fixed at 0.1 to control the level of exploration. For each round of optimization, a batch of $M = 64$ measurements was collected from the physical set-up, which was then reused for $X = 4$ digital update iterations.

To enable operation across all nine visible wavelength channels, we adopted a sequential in-situ training strategy, in which the diffractive layer was optimized for each wavelength in turn. The diffractive layer was first initialized using the parameters obtained from in-silico training, based on the loss functions defined in Equation (10). The subsequent in-situ learning process began at $\lambda_w$ = 710 nm, where the physical hardware was optimized for 20 epochs to fine-tune the diffractive layer parameters for this illumination wavelength. The system was then switched to the next wavelength, $\lambda_w$ = 680 nm. Instead of reinitializing the diffractive layer, the parameters optimized at the previous wavelength were used as the initialization for the next in-situ training step. This sequential optimization procedure was repeated across all illumination wavelengths, from 710 nm to 470 nm. After two complete passes through all nine wavelength channels, an additional 10 epochs of in-situ training were applied to the wavelength channel with the highest loss. The final optimized diffractive layer was then fixed and used for the evaluation across all wavelength channels. This model-free PPO-based in-situ training approach enabled the experimental system to compensate for noise, optical misalignments, calibration errors, and other hardware imperfections that were not fully captured by the numerical model.

**Supporting Information.**

| Name | Description |
|---|---|
| **Supplementary Text** | Training loss function and performance analysis metrics |
| **Video S1** | Video demonstration of wavelength-multiplexed diffractive beam steering |

**REFERENCES**

[1] Z. Li, E. Ahmed, A. M. Eltawil, B. A. Cetiner, *IEEE Trans. Antennas Propag.* **2015**, *63*, 24.

[2] T. Raj, F. H. Hashim, A. B. Huddin, M. F. Ibrahim, A. Hussain, *Electronics* **2020**, *9*, 741.

[3] Y. Monnai, X. Lu, K. Sengupta, *J. Infrared Millim. Terahertz Waves* **2023**, *44*, 169.

[4] S. J. Spector, *J. Opt. Microsyst.* **2022**, *2*, 011003.

[5] R. Morris, C. Jones, M. Nagaraj, *Micromachines* **2021**, *12*, 247.

[6] S. Mansha, P. Moitra, X. Xu, T. W. W. Mass, R. M. Veetil, X. Liang, S.-Q. Li, R. Paniagua-Domínguez, A. I. Kuznetsov, *Light Sci. Appl.* **2022**, *11*, 141.

[7] A. Tuantranont, V. M. Bright, J. Zhang, W. Zhang, J. A. Neff, Y. C. Lee, *Sens. Actuators Phys.* **2001**, *91*, 363.

[8] D. Yang, Y. Liu, Q. Chen, M. Chen, S. Zhan, N. Cheung, H.-Y. Chan, Z. Wang, W. J. Li, *Sci. Rep.* **2023**, *13*, 1540.

[9] D. N. Hutchison, J. Sun, J. K. Doylend, R. Kumar, J. Heck, W. Kim, C. T. Phare, A. Feshali, H. Rong, *Optica* **2016**, *3*, 887.

[10] Y. Liu, C. Zhang, D. M. DeSantis, D. Hu, T. Meissner, J. Notaros, J. Klamkin, *Opt. Express* **2025**, *33*, 7714.

[11] X. Lin, Y. Rivenson, N. T. Yardimci, M. Veli, Y. Luo, M. Jarrahi, A. Ozcan, *Science* **2018**, *361*, 1004.

[12] O. Kulce, D. Mengu, Y. Rivenson, A. Ozcan, *Light Sci. Appl.* **2021**, *10*, 25.

[13] J. Li, D. Mengu, N. T. Yardimci, Y. Luo, X. Li, M. Veli, Y. Rivenson, M. Jarrahi, A. Ozcan, *Sci. Adv.* **2021**, *7*, eabd7690.

[14] E. Goi, S. Schoenhardt, M. Gu, *Nat. Commun.* **2022**, *13*, 7531.

[15] X. Luo, Y. Hu, X. Ou, X. Li, J. Lai, N. Liu, X. Cheng, A. Pan, H. Duan, *Light Sci. Appl.* **2022**, *11*, 158.

[16] C. Liu, Q. Ma, Z. J. Luo, Q. R. Hong, Q. Xiao, H. C. Zhang, L. Miao, W. M. Yu, Q. Cheng, L. Li, T. J. Cui, *Nat. Electron.* **2022**, *5*, 113.

[17] B. Bai, Y. Luo, T. Gan, J. Hu, Y. Li, Y. Zhao, D. Mengu, M. Jarrahi, A. Ozcan, *eLight* **2022**, *2*, 1.

[18] Q. Jia, B. Shi, Y. Zhang, H. Li, X. Li, R. Feng, F. Sun, Y. Cao, J. Wang, C.-W. Qiu, M. Gu, W. Ding, *Optica* **2024**, *11*, 1742.

[19] J. Hu, D. Mengu, D. C. Tzarouchis, B. Edwards, N. Engheta, A. Ozcan, *Nat. Commun.* **2024**, *15*, 1525.

[20] D. Mengu, A. Ozcan, *Adv. Opt. Mater.* **2022**, *10*, 2200281.

[21] C.-Y. Shen, J. Li, Y. Li, T. Gan, L. Bai, M. Jarrahi, A. Ozcan, *Adv. Photonics* **2024**, *6*, 056003.

[22] C.-Y. Shen, J. Li, T. Gan, Y. Li, M. Jarrahi, A. Ozcan, *Nat. Commun.* **2024**, *15*, 4989.

[23] Ç. Işıl, T. Gan, F. O. Ardic, K. Mentesoglu, J. Digani, H. Karaca, H. Chen, J. Li, D. Mengu, M. Jarrahi, K. Akşit, A. Ozcan, *Light Sci. Appl.* **2024**, *13*, 43.

[24] S. Gao, H. Chen, Y. Wang, Z. Duan, H. Zhang, Z. Sun, Y. Shen, X. Lin, *Light Sci. Appl.* **2024**, *13*, 161.

[25] M. Huang, B. Zheng, R. Li, X. Li, Y. Zou, T. Cai, H. Chen, *Laser Photonics Rev.* **2023**, *17*, 2300202.

[26] M. S. S. Rahman, Y. Li, X. Yang, S. Chen, A. Ozcan, *eLight* **2025**, *5*, 32.

[27] A. Chen, Y. Wang, M. S. S. Rahman, Y. Li, A. Ozcan, **2026**, DOI: 10.48550/arXiv.2603.29131.

[28] Y. Wang, A. Chen, M. S. S. Rahman, A. Ozcan, **2026**, DOI: 10.48550/arXiv.2606.01032.

[29] H. H. Zhu, J. Zou, H. Zhang, Y. Z. Shi, S. B. Luo, N. Wang, H. Cai, L. X. Wan, B. Wang, X. D. Jiang, J. Thompson, X. S. Luo, X. H. Zhou, L. M. Xiao, W. Huang, L. Patrick, M. Gu, L. C. Kwek, A. Q. Liu, *Nat. Commun.* **2022**, *13*, 1044.

[30] D. Mengu, A. Tabassum, M. Jarrahi, A. Ozcan, *Light Sci. Appl.* **2023**, *12*, 86.

[31] C.-Y. Shen, J. Li, D. Mengu, A. Ozcan, *Adv. Intell. Syst.* **2023**, 2300300.

[32] Y. Sun, F. Wang, J. Han, G. Qu, Z. Zhang, Y. Wei, C. Yang, Q. Ruan, S. Wang, H. Wei, C. Huang, J. Guan, J. Hu, *Nanophotonics* **2026**, *15*, e70031.

[33] A. Momeni, et al, *Nature* **2025**, *645*, 53.

[34] Y. Li, S. Chen, T. Gong, A. Ozcan, *Light Sci. Appl.* **2026**, *15*, 32.

[35] J. Li, T. Gan, Y. Zhao, B. Bai, C.-Y. Shen, S. Sun, M. Jarrahi, A. Ozcan, *Sci. Adv.* **2023**, *9*, eadg1505.

[36] D. Mengu, Y. Zhao, N. T. Yardimci, Y. Rivenson, M. Jarrahi, A. Ozcan, *Nanophotonics* **2020**, *9*, 4207.

[37] B. Bai, H. Wei, X. Yang, T. Gan, D. Mengu, M. Jarrahi, A. Ozcan, *Adv. Mater.* **2023**, *35*, 2212091.

[38] C.-Y. Shen, P. Batoni, X. Yang, J. Li, K. Liao, J. Stack, J. Gardner, K. Welch, A. Ozcan, *Light Sci. Appl.* **2025**, *14*, 267.

[39] J. Li, Y.-C. Hung, O. Kulce, D. Mengu, A. Ozcan, *Light Sci. Appl.* **2022**, *11*, 153.

[40] Y. Li, J. Li, Y. Zhao, T. Gan, J. Hu, M. Jarrahi, A. Ozcan, *Adv. Mater.* **2023**, *35*, 2303395.

[41] Y. Wang, Y. Li, T. Gan, K. Liao, M. Jarrahi, A. Ozcan, *Nat. Commun.* **2025**, *16*, 5256.

[42] X. Wang, A. Ozcan, **2025**, DOI: 10.48550/arXiv.2512.06658.

[43] N-BK7 | SCHOTT Advanced Optics, https://www.schott.com/shop/advanced-optics/en/Optical-Glass/N-BK7/c/glass-N-BK7, accessed: May, 2023.

[44] J. Li, T. Gan, B. Bai, Y. Luo, M. Jarrahi, A. Ozcan, *Adv. Photonics* **2023**, *5*, 016003.

## FIGURES

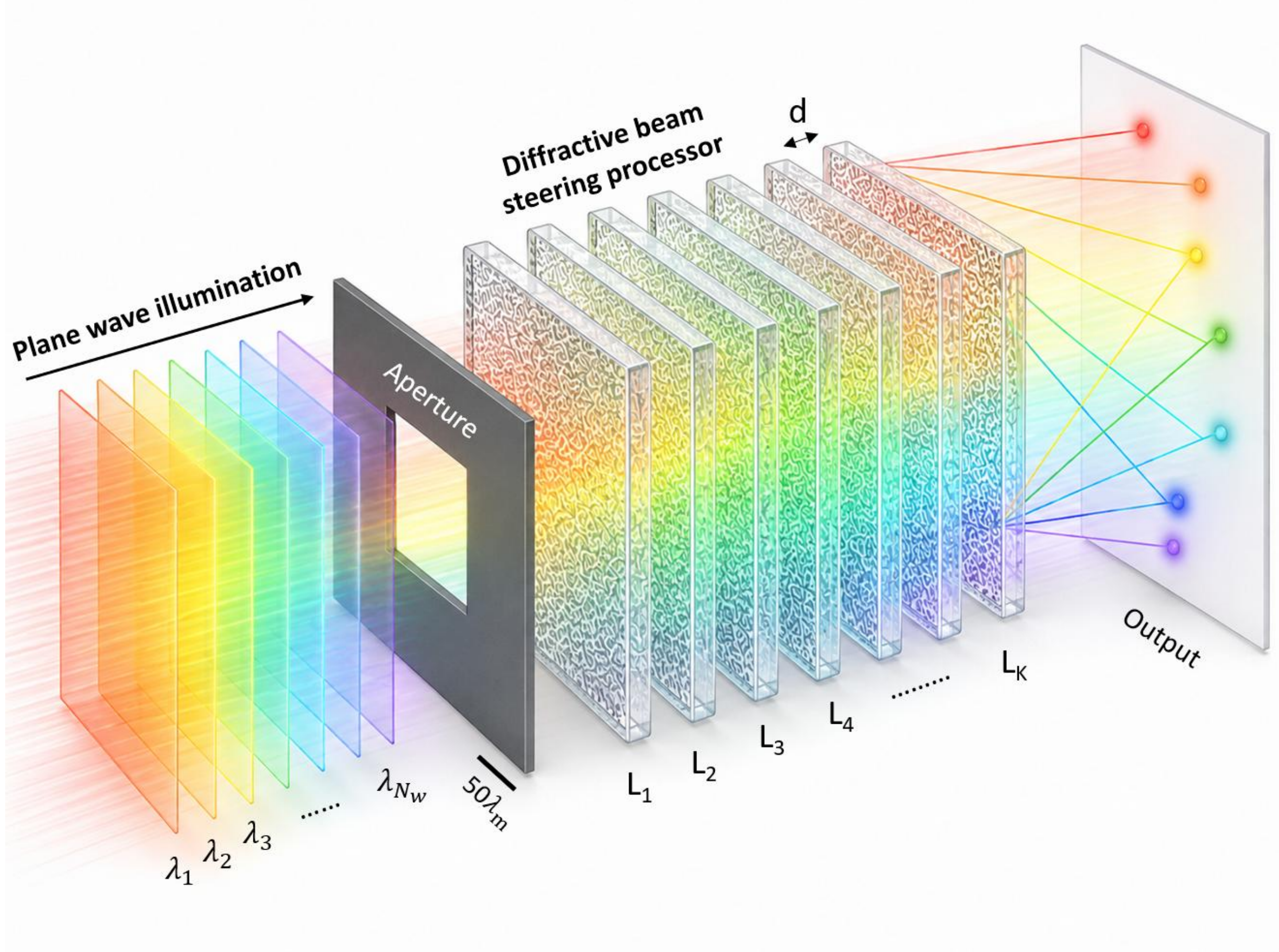

**Figure 1. Schematic of a wavelength-multiplexed diffractive beam steering system.** Illustration of the diffractive beam steering system composed of $K$ cascaded diffractive layers, each containing phase-modulating elements that are jointly optimized using deep learning–based optimization. When illuminated with a set of wavelengths $\{\lambda_1, \lambda_2, \ldots, \lambda_{N_w}\}$, the system projects $N_w$ optical beams toward distinct and programmable angular positions at the output FOV.

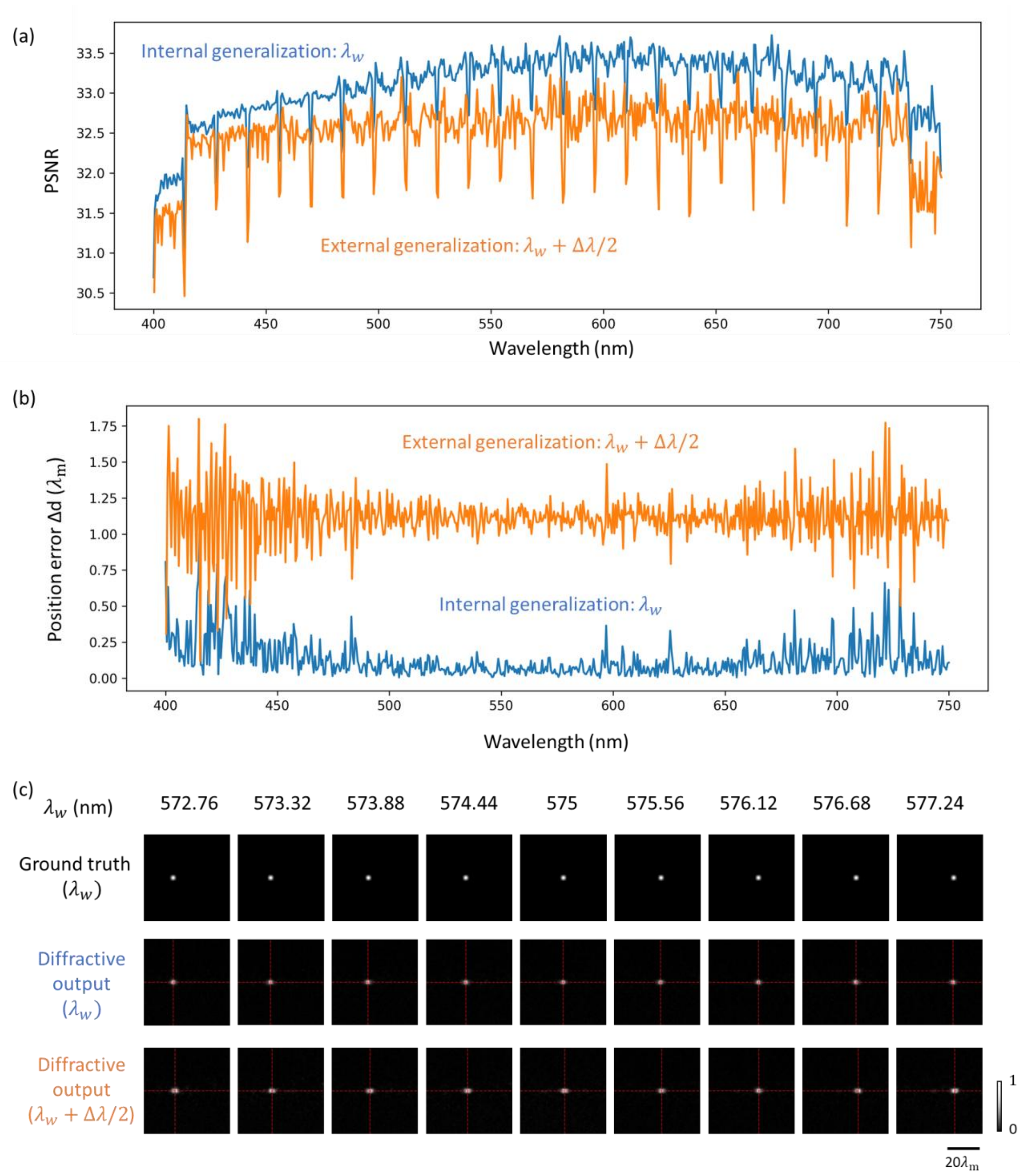

**Figure 2. Performance analysis of the wavelength-multiplexed diffractive beam steering design with $N_w = 625$.** (a) PSNR values of the diffractive beam steering outputs of both internal wavelength channels and external wavelength channels (which were not used during training). Small periodic performance drops in the PSNR curves arise from the 2D zigzag scanning trajectory, where corner points at the turning locations introduce slightly larger errors. (b) Center of mass error values of the diffractive beam steering outputs of both internal wavelength channels and external wavelength channels. (c) Examples of the beam steering outputs, including internal/external wavelength channels. The red dash lines label the center of mass.

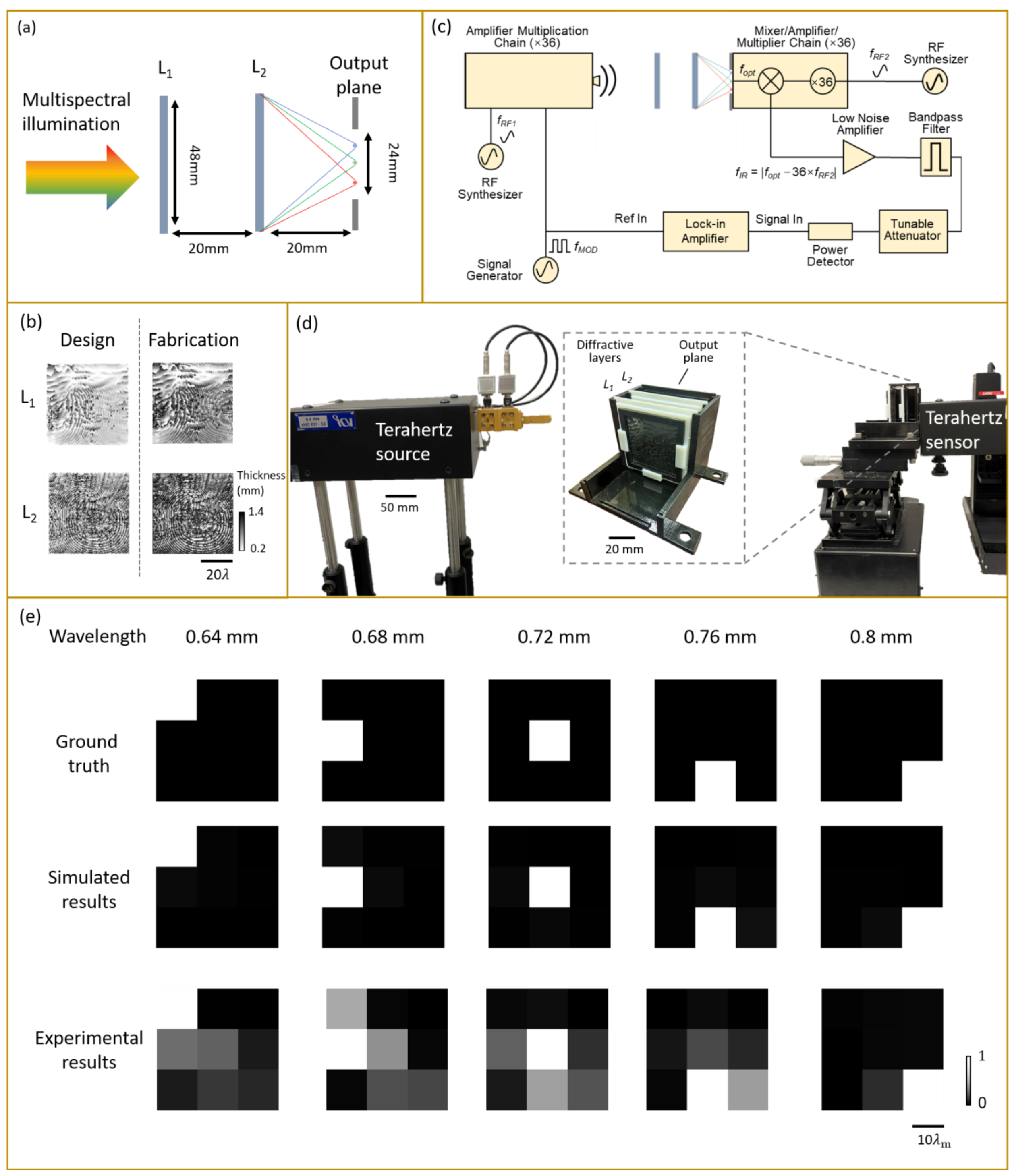

**Figure 3. Experimental demonstration of wavelength-multiplexed diffractive beam steering in the terahertz spectrum.** (a) Illustration of a diffractive beam steering design composed of two diffractive layers ($L_1$, $L_2$). (b) Depiction of the terahertz set-up. (c) Photographs of the experimental set-up, including the 3D-fabricated diffractive beam steering design. (d) Thickness profiles of the optimized diffractive layers (left column) and the photographs of their fabricated versions using

3D printing (right column). (e) Numerically simulated and experimentally measured intensity patterns at the output plane, compared with the ground truth, successfully demonstrating experimental beam steering.

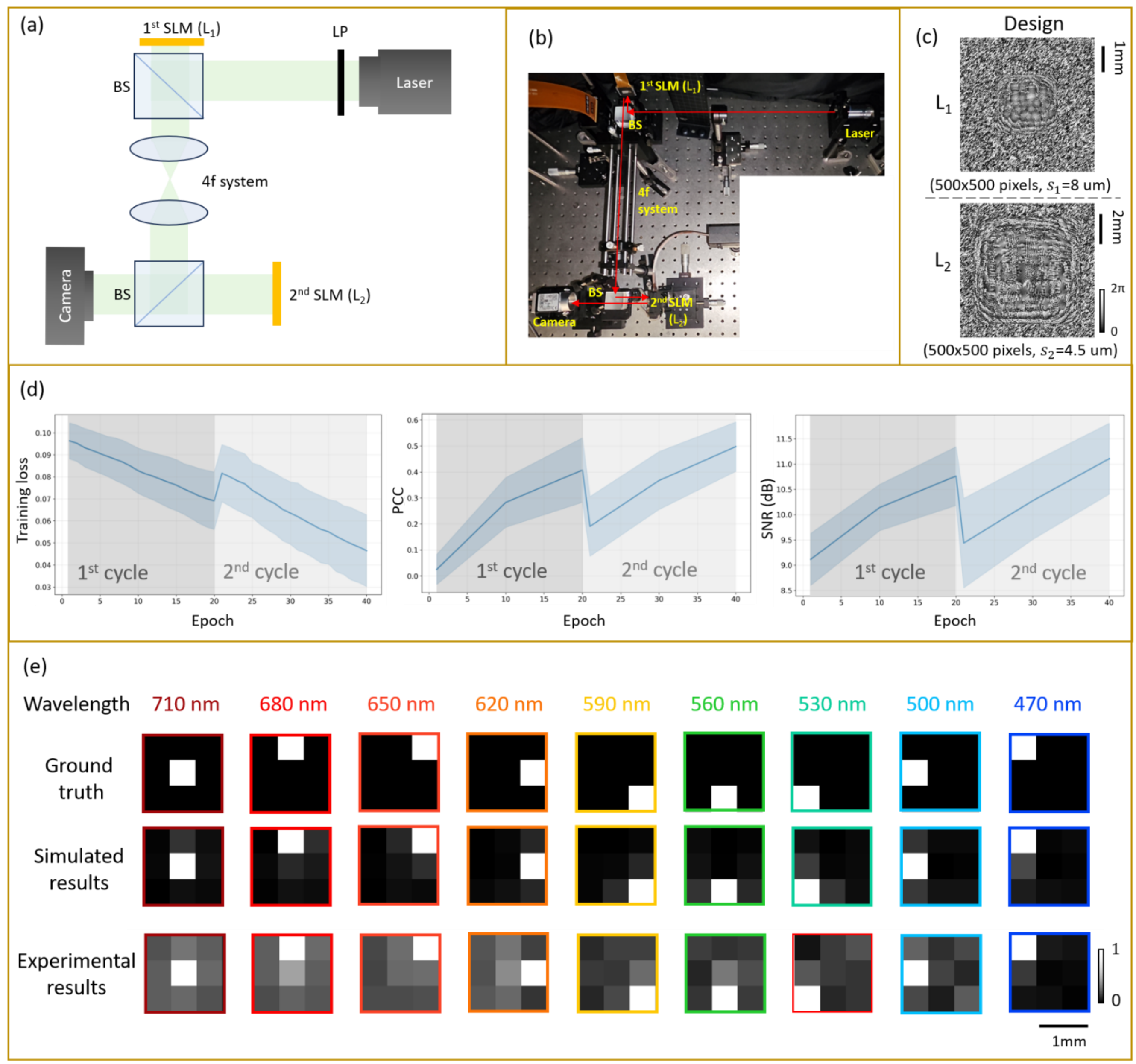

**Figure 4. Experimental demonstration of wavelength-multiplexed diffractive beam steering in the visible spectrum.** (a) Schematic of the diffractive beam steering system, consisting of two cascaded static diffractive layers implemented using two phase-only SLMs. (b) Photograph of the visible-light experimental set-up. (c) Phase profiles of the optimized diffractive layers. (d) Evolution of the average loss, PCC, and SNR values during 40 epochs of in-situ training, calculated between the measured diffractive outputs and the target beam patterns and averaged over nine wavelength channels. The in-situ training consisted of two sequential cycles across all wavelength

channels, with 20 epochs performed for each wavelength in each cycle. The shaded blue regions represent the standard deviation across the nine wavelength channels. (e) Numerically simulated and experimentally measured intensity patterns at the output plane, compared with the ground truth, successfully demonstrating experimental 2D beam steering. The colored boundaries indicate the illumination wavelength associated with each beam steering position.